\documentclass[preprint,aps,pre,amsmath,amssymb,amsfonts]{revtex4}

\def\ra{\rangle}
\def\la{\langle}
\def\bb{\mathbb}
\begin{document}
\baselineskip18pt \thispagestyle{empty}

\begin{center}
\bf    Entanglement property and monogamy relation of generalized
mixed $W$ states
\end{center}
\vskip 1mm

\begin{center}
Ming-Jing Zhao$^{1}$, Shao-Ming Fei$^{1,2}$ and Zhi-Xi Wang$^{1}$

\vspace{2ex}

\begin{minipage}{5in}

\small $~^{1}$ {\small School of Mathematical Sciences, Capital
Normal University, Beijing 100048}

{\small $~^{2}$ Institut f\"ur Angewandte Mathematik, Universit\"at
Bonn, D-53115}


\end{minipage}
\vskip 4mm

(e-mails: zhaomingjingde@126.com, feishm@mail.cnu.edu.cn,
wangzhx@mail.cnu.edu.cn)

\end{center}

\vskip 2mm
\parbox{14cm}
{\footnotesize\quad We have introduced a new class of multipartite
entangled mixed states with pure state decompositions of generalized
$W$ states, similar to Schmidt-correlated states having generalized
GHZ states in the pure state decomposition. The entanglement and
separability properties are studied according to PPT operations.
Monogamy relations related to these states are also investigated.}

\bigskip
{{\it Keywords}: {\footnotesize  generalized mixed $W$ state;
entanglement; monogamy relation}}

\section{Introduction}
Entanglement is a striking feature of quantum systems and
responsible for many quantum tasks such as teleportation, dense
coding, key distribution, error correction etc. \cite{nielsen},
which has provided a strong motivation for the study of detection
and quantification of entanglement. There have been many results
related to separability criterion and entanglement measures whose
effectiveness depends on detailed quantum states. For instance PPT
criterion \cite{Peres A.,M. Horodecki} detects many entangled states
but not bound entangled ones, while the realignment criterion does
\cite{K.Chen}. The entanglement of formation \cite{eof,M.B.P} and
concurrence \cite{con,P.Rungta,S.A} are two well defined
quantitative measures of quantum entanglement. However the
entanglement of formation and concurrence have only explicit
analytical results for some special quantum states such as Werner
states and isotropic states \cite{Terhal-Voll2000,S.M1,S.M2,S.M3,P.
Rungta11}.

For multipartite case there are two well known classes of pure
states, the GHZ and W states. They are shown to be robust against
external flux fluctuations for feasible experimental realizations
\cite{Mun Dae Kim} and the related fidelity can be determined with
an effort increasing only linearly with the number of qubits
\cite{O. Guhne}. Two-party and three-party quantum teleportation
with GHZ state has been discussed. The W state can be also used as
quantum channels for perfect two-party teleportation \cite{A.
Karlsson,W. K. Wootters,P. Agrawal} and quantum key distribution
\cite{J. Joo}. In \cite{Afshin Montakhab} the entanglement dynamics
of GHZ state and W state have been monitored under different models
of system-environment interaction and an exponential decay of
entanglement as a function of time has been obtained. In \cite{B.
Fortescue} a protocol has been presented for distilling maximally
entangled bipartite states between random pairs of parties from
those sharing a tripartite W state. Various experiments have been
set up in the literature for generating three-qubit GHZ and W states
by applying optical systems, nuclear magnetic resonance, cavity QED,
or ion trapping techniques. In \cite{H.H}, they report the scalable
and deterministic generation of four-, five-, six-, seven- and
eight- particle entangled states of $W$ state with trapped ions.

The Schmidt-correlated (SC) states are the mixtures of pure states,
sharing the same Schmidt bases \cite{Rains}. They are generalized to
multipartite case, having the generalized GHZ states as pure state
decompositions in \cite{ming}. An $N$-partite SC state
$\rho_{SC}\in\bb{C}^M\otimes \cdots \otimes \bb{C}^M$ is generally
of the form
\begin{equation}\label{sc}
\rho_{SC} = \sum_{m,n=0 }^{M-1} \rho_{mn} |m \cdots m \rangle \langle n
\cdots n|,
\end{equation}
where $\sum_{m=0}^{M-1} \rho_{mm}=1$. For any pure state
decomposition $\rho_{SC}=\sum_k p_k |\phi_k \rangle\langle \phi_k|$,
$|\phi_k \rangle$ has the from $|\phi_k \rangle =\sum_m \sqrt
{\rho_{mm}} e^{ i \Theta_m^{(k)}} |m \cdots m \rangle $, which is a
kind of generalized $N$-partite $\rm GHZ(N,M)$ state, where
\begin{equation}\label{nmghz}
{\rm GHZ}(N,M) = \frac{1}{\sqrt{M}} (|0 \cdots 0 \rangle + |1 \cdots
1\rangle + \cdots + |M-1, \cdots, M-1 \rangle ).
\end{equation}
An SC state is fully separable if and only if it is PPT with respect
to some subsystems \cite{ming}, and it is either fully separable or
genuine entangled. Here we say an $N$-partite state $\rho$ is fully
separable if it can be written as
\begin{equation}\label{deffullysep}
 \rho=\sum_j p_j \rho^{(j)}_1 \otimes \rho^{(j)}_2 \otimes \cdots \otimes
 \rho^{(j)}_N.
\end{equation}
If $\rho$ is not fully separable, it may be biseparable, which means
it can be written as
\begin{equation}\label{def2sep}
 \rho=\sum_{perm\{i_1, i_2,
\cdots, i_N\}, m, j } p^{(j)}_{i_1, \cdots, i_N,m} \rho^{(j)}_{i_1,
i_2, \cdots, i_m} \otimes \rho^{(j)}_{i_{m+1}, \cdots, i_N},
\end{equation}
where $perm\{i_1, i_2, \cdots, i_N\}$ is a sum over all possible
permutations of the set of indices and $\sum_{perm\{i_1, i_2,
\cdots, i_N\}, m, j} p^{(j)}_{i_1, \cdots, i_N,m}=1$. Here
$\rho^{(j)}_{i_1, i_2, \cdots, i_m}$, $\rho^{(j)}_{i_{m+1}, \cdots,
i_N}$ are density
 matrices associated with the subsystems ${i_1, i_2, \cdots, i_m}$ and
${i_{m+1}, \cdots, i_N}$. We say that $\rho$ is genuine entangled if
it can not be written in the form of (\ref{def2sep}) \cite{O.G}.

In this paper we study
another class of multipartite mixed states, which have the
generalized W states as pure state decompositions.

\section{Generalized mixed W states in multiqubits system }

First we consider multipartite qubit case.
The $N$-partite $|W_N\ra$ state reads,
\begin{equation}\label{wn}
|W_N\ra=\frac{1}{\sqrt{N}}(|0 \cdots 01 \ra + |0 \cdots 10\ra +
\cdots + |1\cdots 00 \ra).
\end{equation}
The $|W_N\ra$ state corresponds to the Dicke state $|D_{1,N}\ra$ and
is therefore an example of a Dicke state. Dicke states have been
investigated already in 1954 by R. H. Dicke, while studying light
emission from a cloud of atoms. In fact they are simultaneous
eigenstates of the collective angular momentum operators
$J_z=\frac{1}{2}\sum_k \sigma_z^{(k)}$ and $J^2$ \cite{O.G}.

The generalized $|W_N\ra$ state is given by,
\begin{equation}\label{gw}
|W_N\ra_g=\sum_{m=1}^Na_m|0\cdots1_m\cdots0\ra=a_1|0\cdots01\ra+a_2|0\cdots10\ra+\cdots+a_N|10\cdots0\ra,
\end{equation}
with $\sum_{m=1}^N|a_m|^2=1$.

For non zero $a_1$, $a_2, \cdots, a_N$, one can show that
$|W_N\ra_g$ is equivalent to $|W_N\ra$ under stochastic local
operation and classical communication (SLOCC) \cite{W. D}, $|W_N
\ra=A_1\otimes A_2\otimes \cdots \otimes A_N|W_N\ra_g$, with
\begin{eqnarray*} A_1= \left(
\begin{array}{cc}
1 & 0  \\
0 & \frac{1}{a_N\sqrt{N}}
\end{array}
\right), ~
A_2= \left(
\begin{array}{cc}
1 & 0  \\
0 & \frac{1}{a_{N-1}\sqrt{N}}
\end{array}
\right),~
 \cdots,~
 A_N= \left(
\begin{array}{cc}
1 & 0  \\
0 & \frac{1}{a_1\sqrt{N}}
\end{array}
\right).
\end{eqnarray*}
Therefore in this case $|W_N\ra_g$ are all genuine $N$-partite entangled states.
Similarly if the number of nonzero coefficients $a_i$ of $|W_N\ra_g$ is $t~(0<t<N)$, then the states are
genuine $t$-partite entangled ones.

Let us consider mixed states with ensembles of pure state
decomposition $\{p_k, |\phi_k\ra\}$, with $|\phi_k \ra =\sum_{m=1}^N
\sqrt{\rho_{mm}}e^{i\theta^{(k)}_m} |0 \cdots 1_m \cdots 0 \ra$,
here $|0\cdots1_m\cdots 0\ra$ denotes a state with the $m$-th
position from right one and others positions zeros, i.e. $|1_N 0
\cdots 0\ra=|10 \cdots 0\ra$, $|01_{(N-1)} \cdots 0\ra=|01 \cdots
0\ra$ and so on. Such states are generally of the form
\begin{equation}\label{rho}
\rho=\sum_{m,n=1}^N \rho_{mn}|0 \cdots 1_m \cdots 0 \ra \la 0 \cdots
1_n \cdots 0|,
\end{equation}
with $\sum_{m=1}^N \rho_{mm} =1$. Such mixed states $\rho$
has only ensemble realizations with
pure states of the form (\ref{gw}).

To study the entanglement property of state (\ref{rho}), we
consider the partial transposition, for instance, with respect to
the $N$-th subsystem, which gives rise to
\begin{eqnarray*} \rho^{PT}= \left(
\begin{array}{ccccccccccc}
0 & 0 & 0 & \rho_{12} & 0 & \cdots & 0 & \cdots & \rho_{1N} & \cdots & 0\\
0 & \rho_{11} & 0 & 0 & 0 & \cdots & 0 & \cdots & 0 & \cdots & 0\\
0 & 0 & \rho_{22} & 0 & \rho_{23} & \cdots & \rho_{2N} & \cdots & 0 & \cdots & 0\\
\rho_{21} & 0 & 0 & 0 & 0 & \cdots & 0 & \cdots & 0 & \cdots & 0\\
0 & 0 & \rho_{32} & 0 & \rho_{33} & \cdots & \rho_{3N} & \cdots & 0 & \cdots & 0\\
\cdots & \cdots & \cdots & \cdots & \cdots & \cdots & \cdots &
\cdots & \cdots & \cdots & \cdots \\
0 & 0 & \rho_{N2} & 0 & \rho_{N3} & \cdots & \rho_{NN} & \cdots & 0 & \cdots & 0\\
\cdots & \cdots & \cdots & \cdots & \cdots & \cdots & \cdots &
\cdots & \cdots & \cdots & \cdots \\
\rho_{N1} & 0 & 0 & 0 & 0 & \cdots & 0 & \cdots & 0 & \cdots & 0\\
\cdots & \cdots & \cdots & \cdots & \cdots & \cdots & \cdots &
\cdots & \cdots & \cdots & \cdots \\
0 & 0 & 0 & 0 & 0 & \cdots & 0 & \cdots & 0 & \cdots & 0\\
\end{array}
\right).
\end{eqnarray*}
By carrying out some elementary transformations, $\rho^{PT}$ can be
transformed into another matrix $(\rho^{PT})^\prime$:
\begin{eqnarray*} (\rho^{PT})^\prime= \left(
\begin{array}{cccc}
A & 0 & 0 & 0 \\
0 & \rho_{11} & 0 & 0 \\
0 & 0 & C & 0 \\
0 & 0 & 0 & D \\
\end{array}
\right),
\end{eqnarray*}
where
\begin{eqnarray*} A= \left(
\begin{array}{cccc}
0 & \rho_{12} & \cdots & \rho_{1N} \\
\rho_{21} & 0 & \cdots & 0 \\
\cdots & \cdots & \cdots & \cdots \\
\rho_{N1} & 0 & \cdots & 0 \\
\end{array}
\right),~~~
 C= \left(
\begin{array}{cccc}
\rho_{22} & \rho_{23} & \cdots & \rho_{2N} \\
\rho_{32} & \rho_{33} & \cdots & \rho_{3N} \\
\cdots & \cdots & \cdots & \cdots \\
\rho_{N2} & \rho_{N3} & \cdots & \rho_{NN} \\
\end{array}
\right),
\end{eqnarray*}
and $D$ is a zero matrix. $C$ is positive semidefinite as $\rho$ is
a density matrix. Therefore the positivity of the matrix
$(\rho^{PT})^\prime$, and hence $(\rho^{PT})$, depends on the matrix
$A$. By deduction we get that the eigenvalues of $A$ are
$\pm(\sum_{j\neq1}|\rho_{1j}|^2)^{\frac{1}{2}}$ and 0. Hence
$\rho^{PT}$ is positive semidefinite if and only if
$(\sum_{j\neq1}|\rho_{1j}|^2)^{\frac{1}{2}}=0$, that is, $\rho_{1j}
=0 $ for $j=2, \cdots, N$. In this case the mixed state (\ref{rho})
becomes
$$
\rho=\rho_{11}|0\cdots0\ra
\la0\cdots0| \otimes |1\ra
\la1|+\sum_{m,n=2}^{N}\rho_{mn}|0\cdots1_m\cdots0\ra
\la0\cdots1_n\cdots0| \otimes |0\ra \la0|.
$$
Therefore it is a bi-separable state with respect to partition $12\cdots N-1$ and $N$ subsystems.
Similar results can be obtained related to partial transpositions with respect to other
subsystems. We have

{\bf Proposition.} The mixed state (\ref{rho}) is a bi-separable one
with respect to partition the $i$-th system and the rest systems if
and only if it is positive semidefinite under partial transposition
with respect to the $i$-th subsystem.

The above property is rather special compared with the ones of SC
states. Moreover, the entanglement of the state $\rho$ is very
robust against particle loss. As the state $|W_N\ra$ remains
entangled even if any $N-2$ parties lose the information, any two
out of $N$ parties possess an entangled state, independent of
whether the remaining $N-2$ parties decide to cooperate with them or
not. Therefore if the mixed state (\ref{rho}) is genuine entangled,
the reduced density matrix of $\rho$, for instance,
$\rho_{\overline{N}}=tr_N
\rho=\rho_{11}|0\cdots0\ra\la0\cdots0|+\sum_{m,n=2}^{N}\rho_{mn}|0\cdots1_m\cdots0\ra
\la0\cdots1_n\cdots0|$ is still a genuine entangled state. The SC
states have no such property. Any kinds of reduced density matrices
of $\rho_{SC}$ states are fully separable. For two dimensional $N$
partite $\rho_{SC}$ state, it is fully separable if and only if
$\rho_{SC}=|\psi\ra$, $|\psi\ra = |0\cdots 0\ra$ or $|\psi\ra =
|1\cdots 1\ra$. But the generalized mixed $W$ state $\rho$
(\ref{rho}) is fully separable if and only if $\rho=\sum_i \rho_{ii}
|0\cdots 1_i \cdots 0\ra \la 0\cdots 1_i \cdots 0 |$, which is
different from SC states.

\section{Monogamy relation of the Generalized mixed W states in multiqubits system}

We now study some monogamy relations related to the generalized
mixed W states in multiqubits system. Recall that the concurrence of
any bipartite pure state $|\psi\ra$ in system
$\mathcal{H}_A\otimes\mathcal{H}_B$ is defined as
$C(|\psi\ra)=\sqrt{2(1-tr\rho_A^2)}$, where
$\rho_A=tr_B|\psi\ra\la\psi|$. The concurrence is then extended to
mixed states $\rho$ by the convex roof:
$C(\rho)\equiv\min_{\{p_i,|\psi_i\ra\}}\sum_ip_iC(|\psi_i\ra)$ for
all possible ensemble realizations
$\rho=\sum_ip_i|\psi_i\ra\la\psi_i|$, where $p_i\geq0$ and
$\sum_ip_i=1$. For a pure three-qubit state $|\psi_{ABC}\ra$,
Coffman, Kundu and Wootters (CKW) \cite{CKW} introduced a monogamy
inequality in terms of concurrence, $C^2_{AB}+C^2_{AC}\leq
C^2_{A(BC)}$, where $C_{AB}$ and $C_{AC}$ are the concurrences of
the mixed states $\rho_{AB}=tr_C(|\psi_{ABC}\ra\la\psi_{ABC}|)$ and
$\rho_{AC}=tr_B(|\psi_{ABC}\ra\la\psi_{ABC}|)$, respectively, and
$C_{A(BC)}$ is the concurrence of $|\psi_{ABC}\ra$ under bipartite
decomposition between subsystems  $A$ and $BC$. The general monogamy
inequality for the case of $n$ qubits is proved in \cite{T. J.
Osborne}. Ref. \cite{Jeong} provided the general monogamy relation
of $|W_N\ra_g$ state with respect to arbitrary partitions. Recently
another monogamy inequality in terms of negativity is deduced in
\cite{Yong}. Negativity is an entanglement measure in two partite
systems which can be expressed as
$N(\rho)=\frac{\|\rho^{PT}\|-1}{2}$, where PT stands for partial
transposition and the trace norm $\|R\|$ is given by
$\|R\|=tr\sqrt{RR^{\dagger}}$. In fact the negativity of state
$\rho$ is essentially the absolute value of the sum of negative
eigenvalues of $\rho^{PT}$. For any three-qubit pure state
$|\psi\ra_{ABC}$, the following CKW-inequality-like monogamy
inequality in terms of negativity holds,
\begin{eqnarray}\label{negabc}
N^2_{AB}+N^2_{AC}\leq N^2_{A(BC)},
\end{eqnarray}
where $N_{AB}$ and $N_{AC}$ are the negativities of the mixed states
$\rho_{AB}$ and $\rho_{AC}$ respectively. $N_{A(BC)}$ is the
negativity of $|\psi_{ABC}\ra$ for the bipartite partition of
subsystems $A$ and $BC$. Similarly one has also
$$
N^2_{BA}+N^2_{BC}\leq N^2_{B(AC)},~~
N^2_{CA}+N^2_{CB}\leq N^2_{C(AB)}.
$$

The general monogamy relation in terms of negativity is given by
\begin{eqnarray}\label{negabcn}
N^2_{A_1A_2}+N^2_{A_1A_3}+\cdots+N^2_{A_1A_N}\leq
N^2_{A_1(A_2A_3\cdots A_N)}.
\end{eqnarray}
Other general monogamy inequalities corresponding to different focused
subsystems $A_i$ can be written down similar to the form (\ref{negabcn}). In the context of quantum
cryptography, such a monogamy property is of fundamental importance
because it quantifies how much information an eavesdropper could
potentially obtain about the secret key extraction. The constraints
on shareability of entanglement lie also at the heart of the success
of many information-theoretic protocols, such as entanglement
distillation \cite{T. J. Osborne}. In this section we prove the
monogamy relation of the mixed state (\ref{rho}) in terms of
negativity.

From the above section we have that the negativity of $\rho$
for a bipartite decomposition between subsystem $A_i$ and the rest subsystems
(non $A_i$) $\overline{A_i}$ is $(\sum_{j\neq
N+1-i}|\rho_{N+1-i,j}|^2)^{\frac{1}{2}}$. Therefore we get
\begin{eqnarray}\label{nega1n}
N_{A_1(A_2\cdots A_N)}^2=\sum_{j\neq N}|\rho_{Nj}|^2.
\end{eqnarray}
As $\rho_{A_1A_2}=tr_{A_3\cdots
A_N}\rho=(\rho_{11}+\cdots+\rho_{N-2,N-2})|00\ra\la00|+\rho_{N-1,N-1}|01\ra\la01|+
\rho_{N-1,N}|01\ra\la10|+\rho_{N,N-1}|10\ra\la01|+\rho_{NN}|10\ra\la10|$,
we get
\begin{eqnarray*}
N_{A_1A_2}=\frac{1}{2}(\sqrt{(\rho_{11}+\cdots+\rho_{N-2,N-2})^2+
4|\rho_{N-1,N}|^2}-(\rho_{11}+\cdots+\rho_{N-2,N-2})).
\end{eqnarray*}

Similarly we can deduce
\begin{eqnarray*}
N_{A_1A_i}&=&\frac{1}{2}(\sqrt{(\rho_{11}+\cdots+\overline{\rho_{N+1-i,N+1-i}}+\cdots+\rho_{N-1,N-1})^2+
4|\rho_{N+1-i,N}|^2}\\
&&-(\rho_{11}+\cdots+\overline{\rho_{N+1-i,N+1-i}}+\cdots+\rho_{N-1,N-1})),
\end{eqnarray*}
where $\overline{\rho_{N+1-i,N+1-i}}$ means that the term is absent in the summation.
As $\sqrt{a+b}\leq\sqrt{a}+\sqrt{b}$, we have
\begin{eqnarray}\label{nega1i}
N_{A_1A_i}&=&\frac{1}{2}(\sqrt{(\rho_{11}+\cdots+\overline{\rho_{N+1-i,N+1-i}}+\cdots+\rho_{N-1,N-1})^2+
4|\rho_{N+1-i,N}|^2}\\\nonumber
&&-(\rho_{11}+\cdots+\overline{\rho_{N+1-i,N+1-i}}+\cdots+\rho_{N-1,N-1}))\\\nonumber
&\leq& |\rho_{N+1-i,N}|.
\end{eqnarray}
From (\ref{nega1n}) and (\ref{nega1i}) we see that (\ref{negabcn})
holds for the mixed state $\rho$. Inequality (\ref{nega1i}) becomes
equality if and only if
$\rho_{11}=\cdots=\overline{\rho_{N+1-i,N+1-i}}=\cdots=\rho_{N-1,N-1}=0$
or $\rho_{N+1-i,N}=0$, $i=2,\cdots, N$. Hence one arrives at that if
the inequality (\ref{negabcn}) becomes equality, then $\rho$ is at
least bipartite $A_i|\overline{A_i}$ separable for some $i$, $1\leq
i\leq N$. In other words, monogamy inequality (\ref{negabcn}) holds
strictly for genuine entangled state (\ref{rho}).

Similarly we have
\begin{eqnarray}\label{nega1a2i}
N_{A_kA_l(A_1\cdots \overline{A_k}\cdots\overline{A_l}\cdots
A_N)}^2=\sum_{i=N+1-k, N+1-l,~j\neq N+1-k,N+1-l}|\rho_{ij}|^2.
\end{eqnarray}
For $A_1A_2|A_3\cdots A_N$ partition, the following equalities
holds:
\begin{eqnarray}
N^2_{A_1A_2(A_3\cdots A_N)}&=&\sum_{i=N-1, N,~j\neq
N-1,N}|\rho_{ij}|^2,\\\label{nega1a2an} N_{A_1A_2(A_k)}
&=&\frac{1}{2}(((\rho_{11}+\cdots+\overline{\rho_{N+1-k,N+1-k}}+\cdots+\rho_{N-2,N-2})^2\\\nonumber
&&+4(|\rho_{N+1-k,N-1}|^2
+|\rho_{N+1-k,N}|^2))^{\frac{1}{2}}\\\nonumber
&&-(\rho_{11}+\cdots+\overline{\rho_{N+1-k,N+1-k}}+\cdots+\rho_{N-2,N-2})).\label{nega1a2ai}
\end{eqnarray}
Therefore one gets
\begin{eqnarray}\label{nega1a2ineq}
N^2_{A_1A_2(A_3)}+\cdots+N^2_{A_1A_2(A_N)}\leq N^2_{A_1A_2(A_3\cdots
A_N)}.
\end{eqnarray}
If inequality (\ref{nega1a2ineq}) becomes an equality, then state
$\rho$ is at least separable under some partition
$A_i|A_j|\overline{A_i}\overline{A_j}$, $1\leq i<j\leq N$, otherwise
it will be a strictly inequality. Generally for any partition
$P_1=\{A_{i_1},\cdots,A_{i_k}\}$,
$P_2=\{A_{i_{k+1}},\cdots,A_{i_{k+l}}\}$, $\cdots$,
$P_s=\{A_{i_{k+s}},\cdots,A_{i_N}\}$, we have
\begin{eqnarray}\label{neganyparti}
N^2_{P_1 P_2}+\cdots+N^2_{P_1 P_s}\leq N^2_{P_1(P_2\cdots P_s)}.
\end{eqnarray}
If this inequality becomes an equality, the state is at least
separable under some partition $A_{t_1}|A_{t_2}|\cdots
|A_{t_k}|\overline{A_{t_1}A_{t_2}\cdots A_{t_k}}$, $1\leq
t_1<t_2<\cdots<t_k\leq N$. On the other hand, we can deduce another
conclusion that mixed state $\rho$ (\ref{rho}) is biseparable if and
only if it is PPT with respect to such partition.

\section{COMMENTS AND CONCLUSIONS}

The results can be generalized to $N$-partite
$d$ dimensional systems. That is
\begin{eqnarray}\label{rho2}
\rho=\sum_{i,k=1}^N\sum_{j,l=0}^{d-1}\rho_{i(j),k(l)}|0\cdots
j_i\cdots0\ra\la0\cdots l_k\cdots0|
\end{eqnarray}
with $\sum_{i=1}^N\sum_{j=0}^{d-1}\rho_{i(j),i(j)}=1$.
We can similarly prove that pure state decomposition of (\ref{rho2}) has the form
$|\psi\ra=\sum_{i=1}^N\sum_{j=0}^{d-1}a_{i(j)}|0\cdots
j_i\cdots0\ra$, which is equivalent to pure state
$|\psi\ra=\sum_{i=1}^N\sum_{j=0}^{d-1}|0\cdots j_i\cdots0\ra$
under SLOCC. Moreover state (\ref{rho2}) is separable with respect to the subsystem $A_i$ $(1\leq
i\leq N)$ if and only if (\ref{rho2}) is PPT with respect to $A_i$.

The monogamy relations can be similarly studied. For example, the
negative eigenvalue of $\rho^{PT}$ with respect to the first subsystem is
$(\sum_{i\neq
N}\sum_{j,l=0}^{d-1}|\rho_{i(j),N(l)}|^2)^{\frac{1}{2}}$. Therefore
the negativity $N^2_{A_1(A_2\cdots A_N)}=\sum_{i\neq
N}\sum_{j,l=0}^{d-1}|\rho_{i(j),N(l)}|^2$. By tedious calculation we
can show that inequality (\ref{negabcn}) and (\ref{neganyparti})
also hold for state (\ref{rho2}). And if the inequalities become
equalities, then corresponding results hold similar to qubit case. While for
SC states, the equalities hold if and only if they are fully
separable, as their reduced matrices are all fully separable.

In summary, similar to SC states having generalized GHZ states in
the pure state decomposition, we have introduced a new class of
multipartite entangled mixed states with pure state decompositions
of generalized $W$ states. The entanglement and separability
properties are studied according to PPT operations. It is shown that
the states are bipartite separable if they are PPT  corresponding
such partition. Monogamy relations related to these states are also
investigated. Although it is still not known if the monogamy
relations in terms of negativity hold for general high dimensional
mixed state, they are true for our class of states. Above all, the
entanglement of these states is very robust against particle loss.
If the mixed state (\ref{rho}) is genuine entangled, the reduced
density matrix of $\rho$ is still a genuine entangled state.

\end{document}